\documentclass{aa} 

\bibpunct{(}{)}{;}{a}{}{,} 

\usepackage{graphicx}
\usepackage{units}
\usepackage[varg]{txfonts}
\begin{document}

\title{qpower2 -- a fast and accurate algorithm for the computation of
exoplanet transit light curves with the power-2 limb-darkening law}

\titlerunning{qpower2}
\author{P.~F.~L.~Maxted \and S.~Gill }           
          
\institute{Astrophysics Group,  Keele University, Keele, Staffordshire,
ST5~5BG, UK\\
\email{p.maxted@keele.ac.uk}
}

\date{Dates to be inserted}

 
  \abstract
{The power-2 law, $I_{\lambda}(\mu) = 1 - c\left(1-\mu^{\alpha}\right)$,
accurately represents the limb-darkening profile for cool stars. It has been
implemented in a few transit models to-date using numerical integration but 
there is as-yet no implementation of the power-2 law in analytic form that is
generally available.}
{Our aim is to derive an analytic approximation that can be used to quickly and
accurately calculate light curves of transiting exoplanets using the power-2
limb-darkening law.}
{An algorithm to implement the power-2 law is derived  using a combination of
an approximation to the required integral and a Taylor expansion of the
power-2 law. The accuracy of stellar and planetary radii derived by fitting
transit light curves with this approximation is tested using light curves
computed by numerical integration of limb-darkening profiles from 3D stellar
model atmospheres.}
{Our algorithm (qpower2) is accurate to about  100\,ppm for broad-band optical
light curves of systems with a star-planet radius ratio $p=0.1$. The
implementation requires less than 40 lines of python code so can run extremely
fast on graphical processing units (GPUs; $\sim$ 1 million models per second
for the analysis of 1000 data points). Least-squares fits to simulated light
curves show that the star and planet radius are recovered to better than  1\%
for $p<0.2$.}
{The qpower2 algorithm can be used to efficiently and accurately analyse large
numbers of high-precision transit light curves using Monte Carlo methods.}
\keywords{binaries: eclipsing -- Techniques: photometric}

\maketitle
%

\section{Introduction}

 Limb darkening is the variation of  specific intensity emitted from a stellar
photosphere as a function of the viewing angle. The advent of very high
precision photometry for transiting exoplanet systems has led to extensive
discussion in the literature of the best way to parameterise limb darkening in
transit models, e.g., \citet{2016MNRAS.457.3573E},
\citet{2013A&A...560A.112M}, \citet{2011MNRAS.418.1165H},
\citet{2008ApJ...686..658S}, \citet{2017AJ....154..111M},
\citet{2017ApJ...845...65N}, \citet{2013MNRAS.435.2152K}, etc. One
well-established result from such studies is that using a linear
limb-darkening law can lead to significant bias in the parameters derived from
the analysis of high quality photometry. For example,
\citet{2016MNRAS.457.3573E} found systematic errors in the radius estimates
for small planets as large as 3\% as a result of using linear limb-darkening
coefficients. There are several alternative ways to parametrize
limb darkening. Among the alternative two-parameter laws used to model the
limb darkening profile $I_{\lambda}(\mu)$ for a given bandpass $\lambda$ is
the quadratic limb-darkening law \citep{1950HarCi.454....1K} --
\[ I_{\lambda}(\mu) = 1- c_1(1-\mu) - c_2(1-\mu)^2, \]
where $\mu$ is the cosine of the angle between the surface normal and the line
of sight. This limb darkening law has the advantage of being relatively simple
and well-understood in terms of the correlations between the coefficients
\citep{2008MNRAS.390..281P, 2011ApJ...730...50K, 2011MNRAS.418.1165H}	 and how
to sample the parameter space to achieve a non-informative prior
\citep{2013MNRAS.435.2152K}. It is frequently used for studies of transiting
exoplanets because several implementations of the algorithm by
\cite{2002ApJ...580L.171M} to rapidly and precisely calculate transit light
curves with quadratic limb darkening are widely available. 
 
Among the limb darkening laws with 2 coefficients, the power-2 limb darkening
law  \citep{1997A&A...327..199H} has been recommended by
\citet{2017AJ....154..111M} as they find that it outperforms other
two-coefficient laws adopted in the exoplanet literature in most cases,
particularly for cool stars. The form of this limb darkening law is
\[I_{\lambda}(\mu) = 1-c\left(1-\mu^{\alpha}\right).\] 
Using  an exponent of $\mu$ rather than a coefficient of some power of
$\mu$ enables this two-parameter law to match accurately the shape of the
limb darkening profile towards the limb of the star using only one extra
parameter cf. a linear limb-darkening law.  The power-2 law has been
implemented in the {\tt ellc} binary star model \citep{2016A&A...591A.111M} and
the {\tt batman} transit model \citep{2015PASP..127.1161K}. For both models,
the transit light curve is calculated using numerical integration. The time
required to perform the numerical integration is not generally a concern if
one is analysing individual targets, but can be a limiting factor if the aim
is to detect and analyse transits in large numbers of high precision light
curves from surveys such as Kepler \citep{2016AJ....152..158T}, K2 \citep{K2}
or TESS \citep{2015JATIS...1a4003R}. \citet{2018A&A...616A..39M} provides a
tabulation of the parameters $c$ and $\alpha$ for cool stars based on
limb darkening profiles calculated using 3-dimensional radiative hydrodynmical
models.  These limb darkening profiles  were tested against the limb darkening
properties of stars measured from Kepler light curves of transiting
exoplanets. The agreement between the computed and observed limb-darkening
parameters was very good for inactive solar-type stars.
 
Here we present the qpower2 algorithm for calculating light curves of
transiting exoplanets and related systems for which the star and planet can be
approximated by spheres and the intensity profile on the star is described by
the power-2 limb darkening law. The algorithm is extremely fast and accurate
enough to model light curves from space-based instruments for systems with a
radius ratio up to $p\approx0.2$. The algorithm can be applied equally to brown
dwarf or low-mass stellar companions to normal stars in eclipsing binary
systems. The deriviation of the algorithm is outlined in section 2. In section
3 we investigate the accuracy of the parameters recovered by least-squares
fitting of transit light curves using the qpower2 algorithm and compare its
performance to the quadratic limb-darkening law. In section 4 we make some
comments regarding the use of the algorithm and execution speed in various
implementations. Our conclusions are given in section 5.

\section{Derivation of the qpower2 algorithm}

 The problem to be solved is to calculate the flux measured by a distant
observer from a spherical star of radius $r_{\star}$ eclipsed by an  opaque
spherical body (``planet'') of radius $r_p \ll r_{\star}$. We set $r_{\star} =
1$ for the following derivation. The star-planet radius ratio is $p =
r_p/r_{\star}$. For this derivation we assume that the smaller companion emits
no flux.

The specific intensity on the stellar disk in some passband $\lambda$ is
described by the power-2 law,
\begin{equation}
  I_{\lambda}(\mu) = I_0\left[1 - c\left( 1 - \mu^\alpha\right) \right]
\end{equation}
where $\mu = \sqrt{1 - r^2}$ is the cosine of the angle between the line of
sight and the normal to the stellar surface, and $r$ is the distance on the sky
from the centre of the star, so the limb of the star is at $r=1$. The
normalizing constant $I_0$ is introduced so that the total flux from the
unocculted star is 1, i.e. 
\begin{equation}
  \int_0^1 I_{\lambda}(r)\,2\pi\,r\,dr = 1,
\end{equation}
where  $I_{\lambda}(r) = I_0\left[ 1 - c + c\left(1 -
r^2\right)^{\gamma}\right]$ and we have
defined $\gamma = \alpha/2$ for convenience. From this definition we obtain
\begin{equation} 
  I_0 = \frac{\alpha+2}{\pi\left[(1-c)\alpha+2\right]}.
\end{equation}

 To calculate the light curve we need to evaluate the integral
\begin{equation}
  F(p,z) = 1 - \int_S I_{\lambda}(r)\,dA,
\end{equation}
where the area to be integrated over, $S,$ is the part of the star obscured
by the planet. In general, evaluating this integral requires use of
hypergeometric functions, which is computational expensive. Instead, we derive
an approximation to this integral by replacing $I_{\lambda}(r)$ by a truncated
Taylor series --
\begin{equation}
  I_{\lambda}(r) \approx I_{\lambda}(r_0) + (r-r_0)I_{\lambda}^{\prime}(r_0) + 
   \nicefrac{1}{2}(r-r_0)^2 I_{\lambda}^{\prime\prime}(r_0) ,
\end{equation}
where  primed symbols denote derivatives with respect to $r$. 

\begin{figure}
  \begin{center}
  \includegraphics[width=0.35\textwidth]{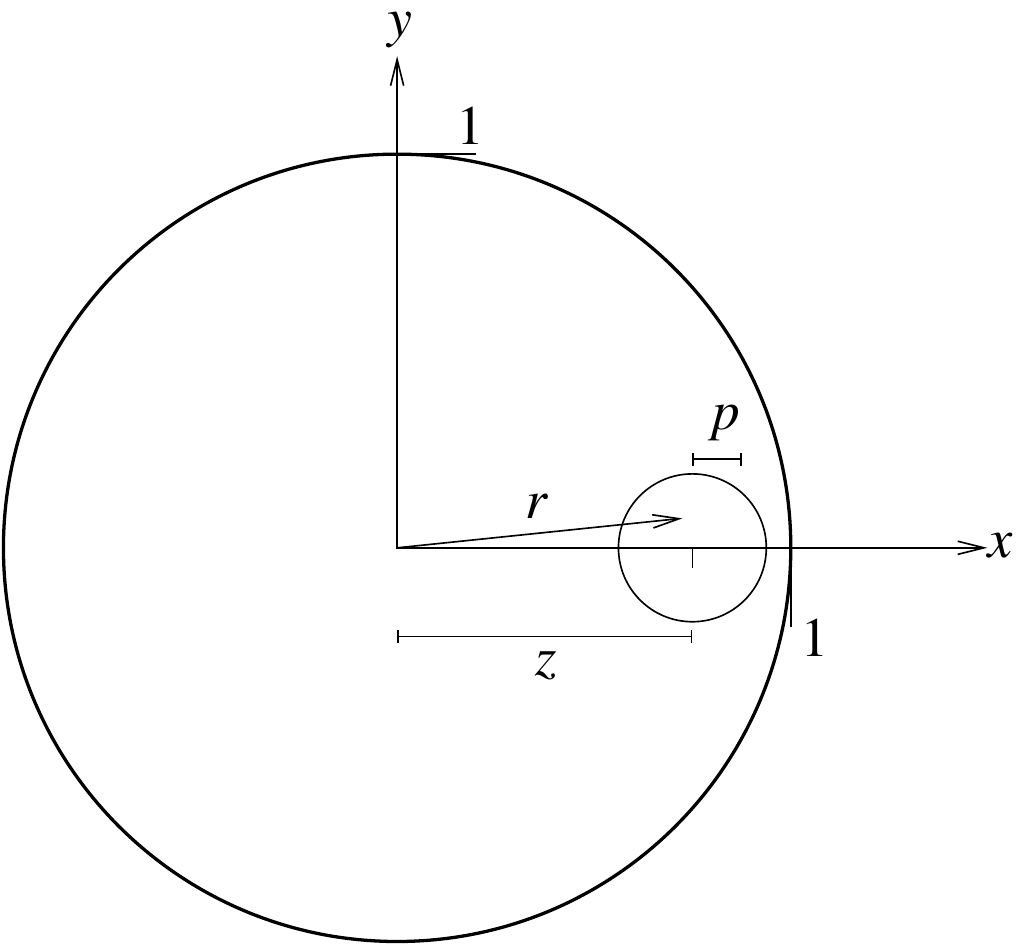}
  \end{center}
  \caption{Coordinate system used for our derivation illustrated for the case
  $|z-1| < p$ . \label{circles}}
\end{figure}

 The coordinate system used for the following derivation is defined such
that the centre of the planet is at the position  $(x,y) = (z,0)$. 
For the case with $z < 1-p$ the disk of the planet lies competely within the
disc of the star, as illustrated in Fig.~\ref{circles}. For these phases we
use $r_0 = z$ as the reference point for the Taylor series expansion. We also
use a Taylor series expansion for $(1-r^2) = (1-x^2-y^2)$ around the value
$y=0$ to obtain the following approximation --
\begin{equation}
  \begin{array}{@{}ll}
    F(p,z) \approx & 1 - 2I_0\int_{z-p}^{z+p}
    \left[1-c+c\left(1-x^2\right)^{\gamma} \right] \sqrt{\ell} \,dx \\
    & +\nicefrac{1}{3}I_0\,\alpha\,c\int_{z-p}^{z+p} 
    \left(1-x^2\right)^{\gamma-1} \ell^{\nicefrac{3}{2}}\,dx,
  \end{array}
\end{equation}
where $\ell = p^2-(z-x)^2$. Approximating $(1-x^2)$ by
$(1-z^2)$ in the second integral and expanding the term in square
brackets in the first integral with a Taylor series around $z$, we obtain
\begin{equation} 
  \label{q1}
  F(p,z) \approx  1 - I_0\,\pi\,p^2\left[c_0 + \nicefrac{1}{4} p^2 c_2 
    - \nicefrac{1}{8} \alpha\,c\, p^2 s^{\gamma-1} \right], 
\end{equation}
where 
\begin{equation}
  c_0 = 1 - c + c\,s^{\gamma},
\end{equation}
\begin{equation}
  c_2 = \nicefrac{1}{2}\alpha\,c\,s^{\gamma-2}\left((\alpha-1)z^2-1\right),
\end{equation}
and $s = 1-z^2$.
 Using $c_0$ only from the term in square brackets in equation (\ref{q1})
is equivalent to using the ``small planet approximation'' described by
\cite{2002ApJ...580L.171M}. 

\begin{figure}
  \begin{center}
  \includegraphics[width=0.4\textwidth]{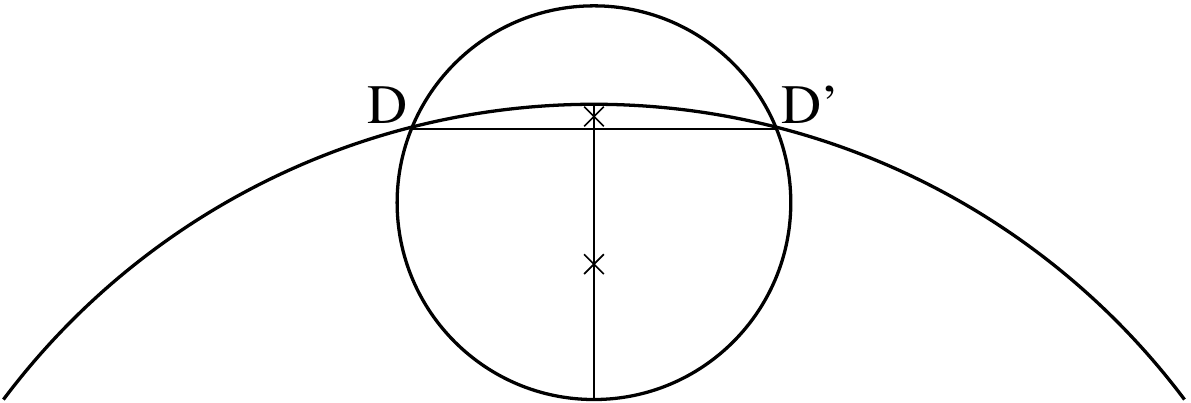}
  \end{center}
  \caption{Geometry of the star and planet at a phase where $1-p < z < 1+p$.
  The chord $DD^{\prime}$ is defined by the intersections between the 
  limb of the star and the limb of the planet. Crosses mark the mid-points of
  the perpendicular bisector of $DD^{\prime}$ between $DD^{\prime}$ and the
  two limbs. These points are at distances $r_a$  and $r_b$ from the origin,
  respectively. \label{intersect}}
\end{figure}

\begin{figure}
  \begin{center}
  \includegraphics[width=0.45\textwidth]{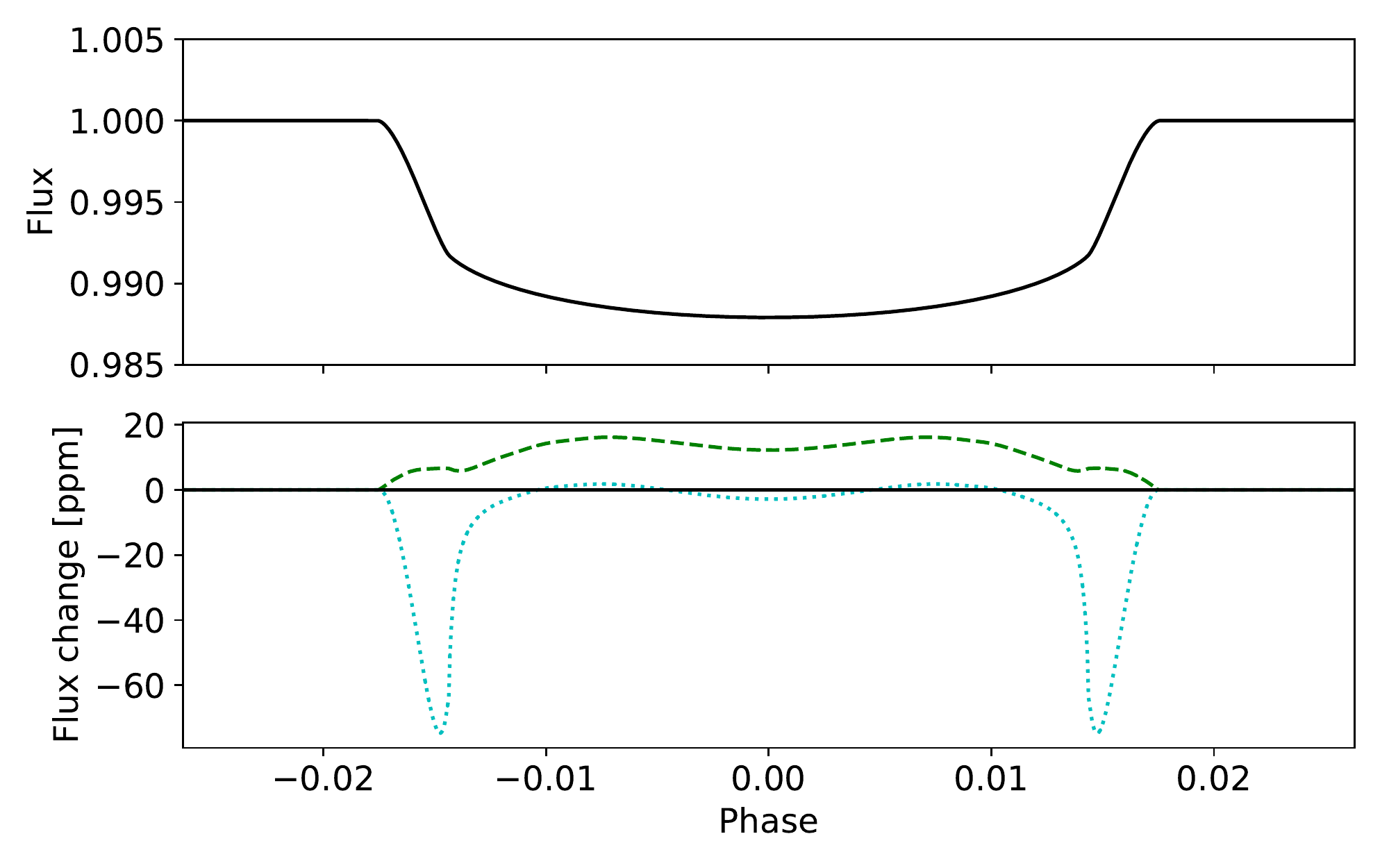}
  \end{center}
  \caption{{\it Upper panel:} light curve computed using the {\tt ellc} light
  curve model  with a limb darkening profile from the STAGGER-grid for
  T$_{\rm eff}=6000$\,K, $\log g=4.5$, ${\rm [Fe/H]} = 0$. The parameters for
  this light curve are $r_{\star} = 0.05$, $p=0.1$ and $b=0$. {\it Lower
  panel:} the difference between light curves computed with the ``sparse''
  numerical grid option in {\tt ellc} and the power-2 limb darkening law
  (dashed green line) or with the qpower2 algorithm (dotted cyan line) and the
  light curve shown in the upper panel. The power-2 and qpower2 light curves
were calculated using the same parameters as in the upper panel.
\label{check_qpower2}} \end{figure}

 A similar approach can be taken for ingress and egress phases of the light
curve where $1-p < z < 1+p$. In these cases the integral is evaluated in two
regions separated by the chord defined by the intersections between the two
limbs. This chord is at a distance $d = \left(z^2 - p^2 +1 \right)/2z$ from
the origin. Care must be taken in choosing the reference point $r_0$ in the
Taylor expansion because $I_{\lambda}^{\prime}(r) \rightarrow \infty$ for
$r\rightarrow 1$. To avoid this problem and to ensure continuity with the
light curve at other phases we choose $r_0 = r_a = (z-p+d)/2$ to evaulate the
integral over the region between the chord and the limb of the planet, and
$r_0 = r_b = (1+d)/2$ for the region between the chord and the limb of the
star. These are the midpoints on the perpendicular bisector of the chord
between the chord and the limb of the star/planet, as illustrated in
Fig.~\ref{intersect}. The region between the chord and the limb of the star is
always small for cases where $p\ll1$ so we only use the first two terms in the
Taylor expansion in this region.

For convenience we define $s_a = 1-r_a^2$, $s_b = 1 - r_b^2$, and also $q =
(z-d)/p$ and $w=\sqrt{p^2 - (d-z)^2}$ before proceeding as before. We then
find that the light curve at these phases can be approximated by
\begin{equation}
  F(z,p) = 1 - I_0\left(J_1 - J_2 + K_1 - K_2\right)
\end{equation}
where
\begin{equation}
  \begin{array}{@{}ll}
    J_1 = &\left[a_0(d-z)-\nicefrac{2}{3}a_1\,w^2 
    +\nicefrac{1}{4}b_2\,(d-z)\left(2(d-z)^2-p^2\right)\right]\,w  \\
       & +  \left(a_0\,p^2 + \nicefrac{1}{4}b_2\,p^4\right)\,\cos^{-1}(q) \\ 
  \end{array}
\end{equation}
\begin{equation}
    J_2 = \alpha\,c\,s_a^{\gamma-1}\,p^4\left(\nicefrac{1}{8}\cos^{-1}(q) +
    \nicefrac{1}{12}\,q\,(q^2-\nicefrac{5}{2})\sqrt{1-q^2}\right), \\
\end{equation}
\begin{equation}
\begin{array}{@{}ll}
    K_1 = & \left(d_0-r_b\,d_1\right)\cos^{-1}(d) \\
    & + \left([r_b\,d+\nicefrac{2}{3}\,(1-d^2)]\,d_1 - 
    d\,d_0\right) \sqrt{1-d^2}\\
 \end{array}
\end{equation}
\begin{equation}
  K_2 = \nicefrac{1}{3}\,c\,\alpha\,s_b^{\gamma+\nicefrac{1}{2}}\,(1-d) 
\end{equation}
The coefficients in these expressions arising from the Taylor expansion of
$I_{\lambda}(z)$ are
\begin{equation}
  b_0 = 1 - c + c\,s_a^{\gamma},
\end{equation}
\begin{equation}
  b_1 =  -\alpha\,c\,r_a\,s_a^{\gamma-1},
\end{equation}
\begin{equation}
  b_2=\nicefrac{1}{2}\alpha\,c\,s_a^{\gamma-2}\left((\alpha-1)r_a^2-1\right),
\end{equation}
\begin{equation}
  d_0 = 1 - c + c\,s_b^{\gamma},
\end{equation}
and
\begin{equation}
  d_1 = -\alpha\,c\,r_b\,s_b^{\gamma-1}.
\end{equation}
These are then grouped according to their common factors to form the
factors
\begin{equation}
  a_0 = b_0 + b_1(z-r_a) + b_2(z-r_a)^2
\end{equation}
and
\begin{equation}
  a_1 = c_1 + 2c_2(z-r_a).
\end{equation}

\begin{figure*}
  \includegraphics[width=0.9\textwidth]{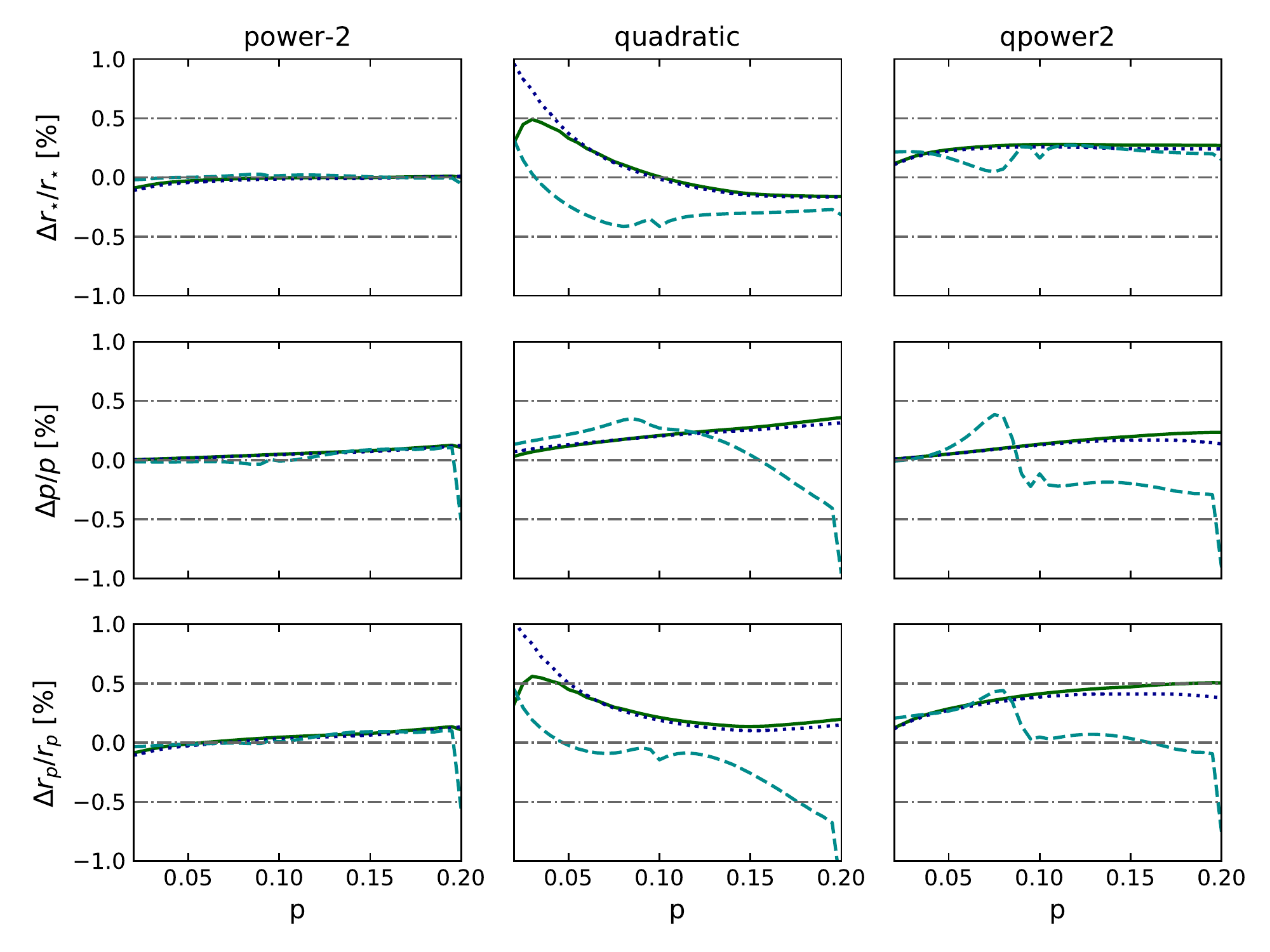}
  \caption{Errors in selected light curve parameters from least-squares fits
  to light curves generated from limb darkening profiles from the 
  STAGGER-grid, as a function of radius ratio, $p$, for three different
  algorithms. From left-to-right: {\tt ellc} light curve model with the
  power-2 limb-darkening law; Mandel \& Agol algorithm for the quadratic
  limb-darkening law; qpower2 algorithm. Results are shown for three
  values of the impact parameter, as follows: $b=0.3$ -- solid green line,
  $b=0.6$ -- dotted blue line, $b=0.9$ -- dashed cyan line. \label{results}}
\end{figure*}

\begin{figure*}
  \includegraphics[width=0.9\textwidth]{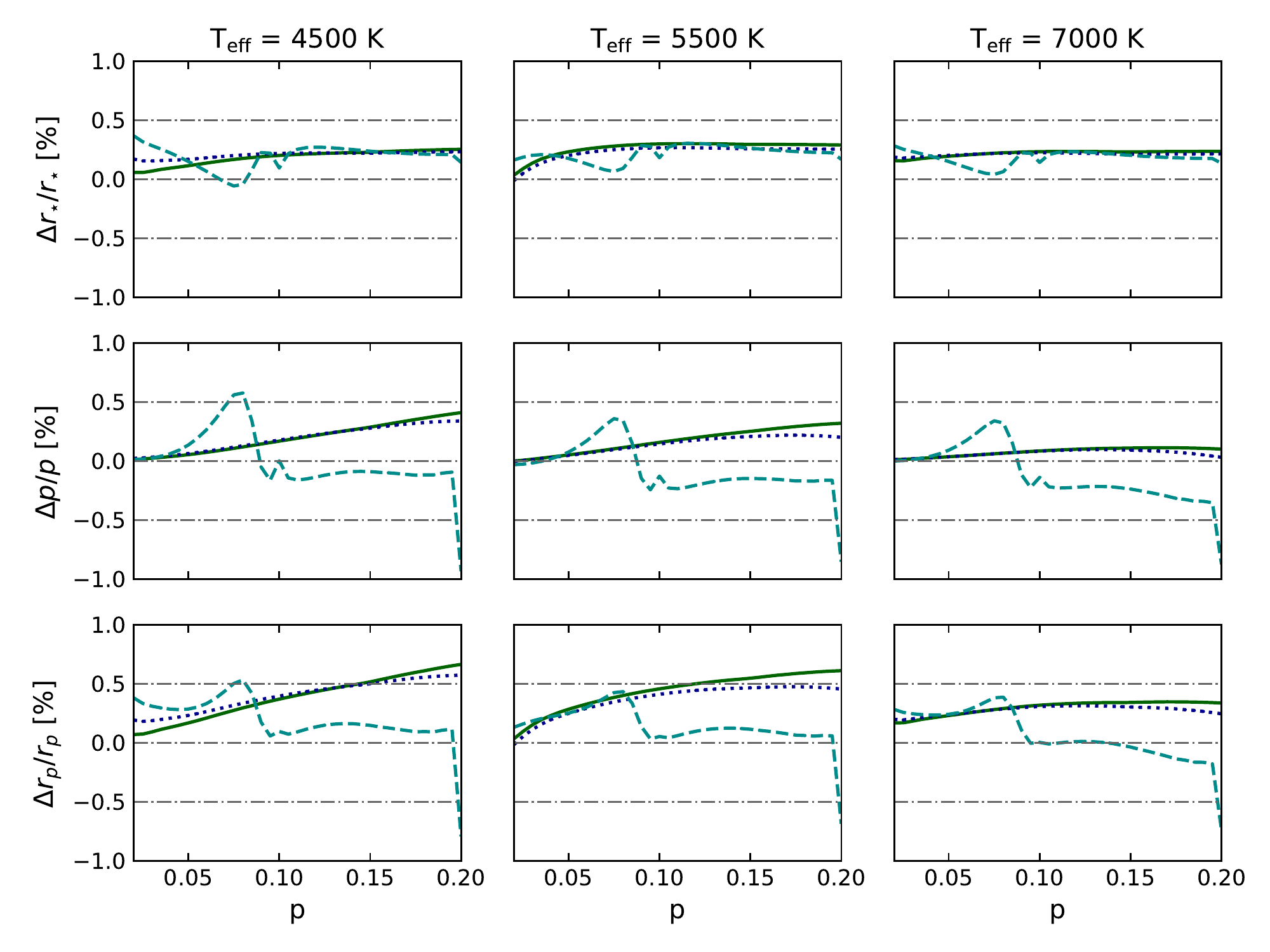}
  \caption{Errors in selected light curve parameters from least-squares fits
  to light curves generated from limb darkening profiles from the 
  STAGGER-grid, as a function of radius ratio, $p$, for  qpower2 algorithm
  for three different values of T$_{\rm eff}$, as noted in the titles to the
  upper panels. Results are shown for three values of the impact parameter, as
  follows: $b=0.3$ -- solid green line, $b=0.6$ -- dotted blue line, $b=0.9$
  -- dashed cyan line. \label{results_Teff}}
\end{figure*}

 An implementation of this algorithm in python is given in the appendix. The
calculations in this paper were done using an  equivalent implementation of
the qpower2 algorithm that is included in the python module
{\tt pycheops}\footnote{\url{https://pypi.python.org/pypi/pycheops/}} that
is currently under development to facilitate analysis of data from the CHEOPS
mission \citep{2017SPIE10563E..1LC}. 

\section{Performance tests}
 In this section we report results of our tests to assess the accuracy of the
qpower2 algorithm. All these tests have been done with light curves calculated
for the CHEOPS passband. The results for the
Kepler  and TESS passbands are very similar. The light curves were simulated
using the {\tt ellc} binary star model assuming that both the star and planet
are spherical and using the ``very\_fine'' numerical integration option so
that numerical noise is no more than a few ppm. The stellar radius has very
little effect on the shape of the transit light curve so we fix this
parameters at a value $r_{\star}/a = 0.1$ for all these simulations. The
lightcurve is simulated for a planet on a circular orbit at 1001 points evenly
distributed over a phase range covering the transit plus 5\% of the transit
width before and after the first and last contact points, respectively.

The limb darkening profile for the star is taken from the tabulated values
provided at 10 values of $\mu$ calculated by \cite{2015A&A...573A..90M} using
the STAGGER-grid 3D stellar atmosphere models. The limb darkening profile is
interpolated to the desired values of T$_{\rm eff}$, $\log g$ and [Fe/H], and
then interpolated onto a regular grid of 101 $\mu$ values using a monotonic
piecewise cubic Hermite interpolating polynomial.\footnote{implemented in the
{\sc python} module \texttt{scipy.interpolate} as the class
\texttt{PchipInterpolator}} We used the values $\log g = 4.5$ and $[{\rm
Fe/H}]=0$ for all the tests presented here. 

For the optimisation of the least-squares fits we used the Nelder-Mead simplex
algorithm as implemented in the {\tt minimize} function of the python package
{\tt scipy.optimize}. We found that this algorithm converged on the correct
solution more reliably than the other algorithms available in this function. 

An example of a simulated light curve is shown in Fig.~\ref{check_qpower2} for
an impact parameter $b=a\cos(i)/r_{\star}=0$ and radius ratio $p =
r_p/r_{\star}=0.1$. This value of $p$ is typical for gas giant planets in
short-period orbits around solar-type stars (``hot Jupiters''). Also shown in
this figure is the light curve calculated using the power-2 limb-darkening law
calculated with {\tt ellc} using values of $c$ and $\alpha$ from
\citep{2018A&A...616A..39M}. From this figure it can be seen that the qpower2
algorithm reproduces light curves for the power-2 limb-darkening law accurate
to better than 0.008\% for these parameters.

\subsection{Accuracy compared to other algorithms}

 We compared the performance of the qpower2 algorithm to numerical integration
of the power-2 limb-darkening law with {\tt ellc} and the algorithm for
quadratic limb darkening by \cite{2002ApJ...580L.171M}. The results are shown
as a function of $p$ in Fig.~\ref{results} for a star with T$_{\rm
eff}=6000$\,K. The results from the power-2 light curve fits are extremely
accurate across the whole range of $p$ for all three values of the impact
parameter used here ($b=0.3, 0.6, 0.9$). Note that we used the ``sparse''
numerical integration grid option in {\tt ellc} to calculate these results.
Better accuracy, if needed, can be achieved using a finer numerical
integration grid but at the expense of increased computation time. 

 The sharp decrease in the accuracy of the recovered values of $r_p$ and
$r_{\star}$ at $p=0.2$ for $b=0.9$ is a result of the eclipse being grazing
for this configuration. The light curve for a grazing eclipse contains very
little information about the geometry of the system so there are large
degeneracies in the least-squares fits and the results are very sensitive to
numerical noise. The eclipse is also very shallow and lacks the characteristic
shape of an eclipse due to a planetary transit that is typically used to
identify these systems in photometric surveys so we ignore grazing eclipses
for the remainder of this discussion.

 The radii and radius ratio determined with the quadratic limb-darkening law
are accurate to approximately 0.5\% over the same range of $p$ and $b$. The
performance of this algorithm in terms of the recovered values of $r_p$ and
$r_{\star}$ is worst for small values of $p$, while the recovered value of $p$
is accurate to better than 0.5\% for all values of $p$ for impact parameters
$b=0.3$ and $b=0.6$. We have not investigated these trends in detail but
strongly suspect that they are due to the poor match between the quadratic
limb-darkening law and the realistic limb darkening profiles for solar-type
stars from the STAGGER-grid at small values of $\mu$, i.e. towards the limb of
the star. 

The performance of the qpower2 algorithm overall is very similar to the
quadratic limb-darkening algorithm, i.e., the results are accurate to better
than approximately 0.5\% for transits with $p<0.2$. One clear
difference is that the qpower2 algorithm performs better than quadratic
limb-darkening law for $p\la 0.06$. For $b=0.3$ and $b=0.6$ there is a small
bias in the recovered values that varies slowly with $p$. This suggests that
it should be possible to correct for this bias in the analysis of high-quality
light curves using simulations similar to those presented here. The best fit
from the least-squares fit using the qpower2 algorithm can also be used as an
accurate starting point for further least-squares fits using numerical
integration of the power-2 limb darkening law with {\tt ellc} or {\tt batman}.

\subsection{Accuracy as a function of effective temperature}

 The results as a function of stellar effective temperature,  T$_{\rm eff}$,
for the qpower2 algorithm with $p=0.1$ are shown in
Fig.~\ref{results_Teff}. The range T$_{\rm eff} = 4500$\,K to 7000\,K is set
by the range of stellar effective temperature available from the STAGGER-grid
for $\log g = 4.5$. The accuracy of the recovered parameter values is quite
consistent across the full range of T$_{\rm eff}$ and is, in general, better
than 0.5\%.

\section{Implementation notes and timing tests}

 In this section we make some comments regarding the implementation of the
algorithm and present the results of some tests we have conducted to assess
the speed of the qpower2 algorithm.

\subsection{Python implementation}
 The python implementation of the qpower2 algorithm shown in Fig.~\ref{python}
uses the function {\tt select} to assign the output of either function {\tt
q1} or {\tt q2} to the output array depending on whether $z\le p$ or
$|z-1|<p$. This requires that these functions are valid for any input value of
$z$. We have used the {\tt clip} function to restrict the value of $z$ and so
avoid invalid calculations inside these functions. Similarly, the function
{\tt finfo(0.0).eps} is used to generate a small floating-point number
(typically $2^{-52}$) to avoid errors due to an attempt to raise 0 to a
negative power.  Alternative methods for applying the conditions $z\le p$ and
$|z-1|<p$ can avoid some of these complications and may be faster in some
cases since fewer calls to {\tt q1} and  {\tt q2} will be required. For
clarity, we have not included good programming practices such as  error and
warning message generation, checks for invalid  input parameters, in-line
documentation or  comments  in this code fragment. 

 The implementation of qpower2 in the current development version of 
{\tt pycheops} (0.0.13) uses a loop to pass once through the input values of $z$
with an  {\tt if \dots then \dots else if \dots } logical structure to apply
the conditions $z\le p$ and $|z-1|<p$. This structure is well suited to
``just-in-time'' compilation and optimisation using the package
{\tt numba}.\footnote{https://numba.pydata.org/} We found this to be an
effective and easy way to dramatically improve the speed of the calculation,
as described below. 

\subsection{Comparison with other algorithms}

\begin{table*}
\caption{The execution time per light curve for simulations containing either
1000 data points (Data set A) or 3840 data points (Data set B) for the transit
of a star by a planet with $p\approx0.1$. The notes for batman give the value
of the {\tt max\_err} option that is used to set the number of integration
steps such that the maximum error due to numerical noise does not exceed the
value given in ppm.}
\label{speedtest}
\begin{center}
\begin{tabular}{@{}lllrl}
\hline
  Algorithm & Limb darkening & Processor & 
  \multicolumn{1}{l}{Execution } &Notes \\
  &    &           & \multicolumn{1}{c}{time [$\mu$s] }  \\
\hline
\multicolumn{5}{@{}l}{\bf Data set A}\\
  qpower2, {\tt pycheops} &  power-2   & 2\,GHz CPU &  113 & Optimised using {\tt
numba} \\
  qpower2, {\tt pycheops} &  power-2   & 2\,GHz CPU & 5550 & No optimisation \\
  qpower2, Fig. A.1          &  power-2   & 2\,GHz CPU &  570 \\
  batman &  quadratic & 2\,GHz CPU &  169 \\
  batman&  power-2   & 2\,GHz CPU & 2380 & maxerr=1 (default)\\
  batman&  power-2   & 2\,GHz CPU &  247 & maxerr=70 \\
\noalign{\smallskip}
\multicolumn{5}{@{}l}{\bf Data set B}\\
qpower2, C/OpenMP& power-2& 4.8\,GHz CPU &  357 & \\
qpower2, C/OpenMP& power-2& 4.8\,GHz CPU $\times 8$&  112  \\
qpower2, C/OpenMP& power-2& GTX1080 GPU& 13.2 & Return array of models \\
qpower2, C/OpenMP& power-2& GTX1080 GPU& 2.5  & Return 
 $\log {\mathcal {L}}$  values only\\
  \noalign{\smallskip}
\hline
\end{tabular}
\end{center}
\end{table*}

 We tested the speed of various algorithms to calculate the transit light
curve of a system with $p=0.1$, $r_{\star}=0.1$ and $b=0$. For all the
algorithms tested we simulated a light curve with 1000 observations uniformly
sampled over one transit plus 5\% in phase before and after the start and end
of the transit.  

 The algorithms tested were: the qpower2 implementation from {\tt pycheops} with and
without optimisation using {\tt numba}; the qpower2 implementation from
Fig.~\ref{python}; {\tt batman} using quadratic limb darkening; {\tt
batman} using power-2 limb-darkening. Quadratic limb darkening in {\tt batman}
uses a variant of the algorithm by \citet{2002ApJ...580L.171M}. The power-2
algorithm in  {\tt batman}  uses a numerical integration scheme with the
option to set the maximum numerical error in ppm. We ran simulations with the
default option ${\tt maxerr = 1}$ and also simulations with ${\tt maxerr =
70}$ for direct comparison with the qpower2 algorithm. These tests were all
performed on an Apple MacBook Pro with a 2\,GHz Intel\textregistered\ Core i7
CPU.  Timings were calculated using the {\tt \%timeit} function in IPython. 

From the results shown in Table~\ref{speedtest} for these simulations (Data
set A) we see that the optimised qpower2 implementation from {\tt pycheops} is the
fastest of the algorithms tested and is just over twice as fast as the {\tt
batman} algorithm for power-2 limb-darkening with ${\tt maxerr = 70}$, and
50\% faster than {\tt batman} with quadratic limb darkening.

\subsection{CPU versus GPU timing tests}

The qpower2 algorithm is a small piece of code than can be executed
in parallel on a data set of moderate size. This makes it well-suited to
acceleration by executing it on a graphical processing unit (GPU). We have
experimented with this option using the CUDA\textregistered\ toolkit by
NVIDIA\textregistered.\footnote{https://developer.nvidia.com/cuda-toolkit}
This option works particularly well if the code can be refactored to 
send a single array with multiple sets of parameter values to the GPU, i.e. it
is much faster to loop over the calls to the qpower2 function on the GPU
rather than the CPU. Another effective optimisation for Markov chain
Monte Carlo routines such as {\sc emcee} \citep{2013PASP..125..306F} is to
calculate the log-likelihood ($\log {\mathcal {L}}$) for each model on the GPU
and to return only these values, rather than incurring the overhead of
returning the simulated light curve from the GPU to the CPU.

 We used a PC running linux (Ubuntu 17.10) with eight 4.2\,GHz
Intel\textregistered\  Core i7-7700K CPUs (overclocked to 4.8 GHz) and a
GeForce GTX 1080 GPU to compare the execution speed of the qpower2 algorithm
running on CPUs and GPUs. The parameters used for these simulations were the
similar  to those for the comparison between algorithms in the previous
section except that we used 3840 points per eclipse. We used an implementation
of the qpower2 algorithm written in C using
OpenMP\footnote{http://www.openmp.org} for parallelization. The results
are shown in Table~\ref{speedtest} (Data set B).

 From Table~\ref{speedtest} we see that the speed-up using this option does
not scale with the number of CPUs used. This is a consequence of the overheads
in the parallelization. Nevertheless, with 8 CPUs it is possible to increase
the speed of an MCMC analysis by more than a factor of 3. However, gains in
speed of an order of magnitude  are possible by using a GPU to calculate the
light curves, and an additional speed-up by more than a factor of $\times 5$
is possible if the MCMC code can refactored so that the log-likelihood ($\log
{\mathcal {L}}$) calculation is performed on the GPU, rather than returning
the computed light curve to the CPU.  With these optimisations it is possible
to calculate up to 1 million $\log {\mathcal {L}}$ values per second on a GPU
for a transit light curve with 1000 data points.

\section{Conclusion}
 The qpower2 algorithm is straightforward to implement, very fast and
sufficiently accurate to model the light curves of transiting exoplanet
systems and related objects from instruments such as Kepler, TESS, CHEOPS and
PLATO \citep{2014ExA....38..249R}. For a typical hot Jupiter system, the
log-likelihood for a transit light curve of 1000 observations can be computed
for a model light curve accurate to 100\,ppm in approximately 1$\mu$s on a
GPU. This makes this algorithm an attractive choice for {\it en masse}
analysis of light curves from these massive photometric surveys. 

\begin{acknowledgements}
 SG acknowledges support from Doctoral Training Partnership grant number
  ST/N504348/1 from the Science and Technology Facilities Council (STFC). PM
  acknowledges support from STFC research grant number ST/M001040/1.
\end{acknowledgements}

\bibliographystyle{aa} 
\bibliography{mybib}

\appendix
\section{Python implementation}
An implementation of the qpower2 algorithm written for python version 3.6
is shown in Fig.~\ref{python}. Note that this implementation is written for
clarity rather than optimised for speed. 

\begin{figure*}
\begin{verbatim}
def qpower2(z,p,c,alpha):
  from numpy import arccos, sqrt, pi, clip, select, finfo
  I_0 = (alpha+2)/(pi*(alpha-c*alpha+2))
  g = 0.5*alpha
  def q1(z,p,c,alpha):
      zt = clip(abs(z), 0,1-p)
      s = 1-zt**2
      c0 = (1-c+c*s**g)
      c2 = 0.5*alpha*c*s**(g-2)*((alpha-1)*zt**2-1)
      return 1-I_0*pi*p**2*(c0 + 0.25*p**2*c2 - 0.125*alpha*c*p**2*s**(g-1))
  
  def q2(z,p,c,alpha):
      zt = clip(abs(z), 1-p,1+p)
      d = clip((zt**2 - p**2 + 1)/(2*zt),0,1)
      ra = 0.5*(zt-p+d)
      rb = 0.5*(1+d)
      sa = clip(1-ra**2,finfo(0.0).eps,1)
      sb = clip(1-rb**2,finfo(0.0).eps,1)
      q = clip((zt-d)/p,-1,1)
      w2 = p**2-(d-zt)**2
      w = sqrt(clip(w2,finfo(0.0).eps,1))
      b0 = 1 - c + c*sa**g
      b1 = -alpha*c*ra*sa**(g-1)
      b2 = 0.5*alpha*c*sa**(g-2)*((alpha-1)*ra**2-1)
      a0 = b0 + b1*(zt-ra) + b2*(zt-ra)**2
      a1 = b1+2*b2*(zt-ra)
      aq = arccos(q)
      J1 = ( (a0*(d-zt)-(2/3)*a1*w2 + 0.25*b2*(d-zt)*(2*(d-zt)**2-p**2))*w
              + (a0*p**2 + 0.25*b2*p**4)*aq )
      J2 = alpha*c*sa**(g-1)*p**4*(0.125*aq +
              (1/12)*q*(q**2-2.5)*sqrt(clip(1-q**2,0,1)) )
      d0 = 1 - c + c*sb**g
      d1 = -alpha*c*rb*sb**(g-1)
      K1 = ((d0-rb*d1)*arccos(d) +
              ((rb*d+(2/3)*(1-d**2))*d1 - d*d0)*sqrt(clip(1-d**2,0,1)) )
      K2 = (1/3)*c*alpha*sb**(g+0.5)*(1-d)
      return 1 - I_0*(J1 - J2 + K1 - K2)
  
  return select( [z <= (1-p), abs(z-1) < p],  
      [q1(z, p, c, alpha), q2(z, p, c, alpha)], default=1)
\end{verbatim}
  \caption{A python implementation of the qpower2 algorithm. 
  \label{python}}
\end{figure*}

\end{document}